\documentclass[manuscript,screen]{acmart}
\setcopyright{none}
\renewcommand\footnotetextcopyrightpermission[1]{}

\AtBeginDocument{%
  }

\begin{document}

\title{Constraint-Driven Context Engineering: Designing Domain Interfaces for AI Systems}

\author{Xiwei Xu}
\correspondingauthor
\email{xiwei.xu@csiro.au}
\orcid{1234-5678-9012}
\author{Chen Wang}
\email{chen.wang@csiro.au}
\author{Mengmeng Yang}
\email{mengmeng.yang@csiro.au}
\author{Yipeng Zhang}
\email{yipeng.zhang@csiro.au}
\author{Jacky Jiang}
\email{jacky.jiang@csiro.au}
\author{Suyu Ma}
\email{suyu.ma@csiro.au}
\author{Youyang Qu}
\email{youyang.qu@csiro.au}
\author{Ming Ding}
\email{ming.ding@csiro.au}
\author{Liming Zhu}
\email{liming.zhu@csiro.au}
\affiliation{%
  \institution{CSIRO}
  \city{Sydney}
  \country{Australia}
}

\renewcommand{\shortauthors}{Xiwei et al.}

\begin{abstract}

Generative AI systems are increasingly deployed to address domain problems. These systems operate under technical, regulatory, institutional, and normative constraints that define acceptable AI behaviour and outcomes within their domains. We observe a recurring pattern in our industry engagement: partners often arrive with a functioning but relatively generic AI solution. The challenge is no longer to build an AI system from scratch, but to improve the quality and domain appropriateness of an AI-generated solution. In these settings, the limiting factor is often the quality, scope, and structure of the context available to the system. Yet, existing context engineering approaches primarily focus on supplying domain knowledge through retrieval, memory, and tools, with limited support for systematically identifying and operationalising the constraints that govern AI systems in their operational environments.

This paper proposes Constraint-Driven Context Engineering (CDCE), a design approach for engineering domain interfaces for AI systems. Drawing on software architecture design and Domain-Driven Design (DDD), CDCE treats domain constraints as first-class design drivers. It identifies and characterises constraints, determines the required context assets, and designs representations through which these assets are made available to AI systems.

We conducted a comparative multiple-case study with industry and public-sector partners across educational assessment, healthcare decision support, and financial-distress prediction. Depending on their characteristics, constraints can guide AI behaviour, enforce permissible boundaries, or support verification of AI-generated outcomes. The cases demonstrate CDCE's applicability across contrasting domains and show how constraint characteristics shape the resulting domain interfaces.

\end{abstract}


\maketitle

\section{Introduction}

Large language models (LLMs) are increasingly deployed across domains to assist human experts by interpreting information, generating recommendations, and supporting complex decisions~\cite{brown2020language, wei2022chain}. Unlike widely adopted general-purpose conversational applications, these AI systems operate within domain-specific environments where AI behaviour is governed by domain knowledge as well as technical, regulatory, institutional and normative constraints in the domain.

Understanding the domain has long been recognised as a fundamental principle of software design. Domain-Driven Design (DDD) advocates that effective software should be built around a deep understanding of the problem domain, where domain experts, domain models, and ubiquitous language guide the design process~\cite{evans2004domain}. Software architecture design complements this perspective by treating functional requirements, quality attributes, and constraints imposed by technology and deployment environments as design drivers. Constraints are recognised as one of the primary architectural drivers~\cite{Perry1992}, alongside functional requirements and quality attributes~\cite{bass2021software}.

A similar challenge is emerging in AI systems. Existing context engineering approaches often focus on making external knowledge available to generative AI through retrieval, memory, tools, and agentic workflows~\cite{lewis2020retrieval, yao2022react, schick2023toolformer}. However, making domain knowledge available alone is insufficient for AI systems operating in domain-specific settings. AI behaviour is also governed by constraints that determine what constitutes acceptable decisions, appropriate actions, and valid justifications. Some of these constraints are explicit, such as regulations, policies, guidelines, and technical requirements, but may be fragmented across heterogeneous documents, systems, and organisational processes. Others are implicit, embedded in established practices, professional judgement, and human knowledge. Consequently, domain constraints do not naturally exist in a form that AI systems can directly access and consistently apply. Explicit constraints need to be systematically structured and operationalised, while implicit constraints additionally need to be elicited and externalised before they can be incorporated into AI behaviour and decision-making.

We conceptualise context engineering as designing a domain interface between an AI system and the domain in which it operates. The domain interface makes relevant domain constraints available in forms suitable for AI systems. It structures explicit constraints, externalises implicit constraints, and operationalises constraints as machine-accessible context for guiding behaviour, enforcing permissible boundaries, and verifying AI-generated outcomes. In this way, we move context engineering beyond a knowledge-centric perspective towards a constraint-driven perspective. We propose Constraint-Driven Context Engineering (CDCE), a design approach for systematically identifying, representing, and operationalising the constraints that govern AI behaviour within a domain. The approach treats constraints as first-class architectural drivers for determining what context is required, what context assets need to be constructed or curated, and how these assets should be represented and made available to AI systems.

We conducted a comparative multiple-case study across educational assessment, healthcare decision support, and financial-distress prediction. The case study examines how domain constraints are translated into context assets, context representations, and how human review is incorporated. 
Together, the cases demonstrate how constraint-driven design can make 
relevant domain conditions explicit, structured, and available to AI systems for different purposes. 

The contributions of this paper are as follows:

\begin{itemize}
    \item We propose Constraint-Driven Context Engineering, a design approach for systematically engineering domain constraints as a domain interface that mediates between an AI system and its operational domain.
    \item We conceptualise domain constraints as first-class design drivers that guide, restrict, and evaluate AI behaviour within specific domains.
    \item We empirically examine the applicability of the proposed CDCE through a comparative multiple-case study spanning educational assessment, healthcare decision support, and financial-distress prediction.
\end{itemize}

\section{Related Work}

\subsection{Context Engineering for LLM Systems}

Context engineering has emerged as an approach to systematically constructing and managing the information fed to LLMs during inference. Its foundations can be traced back to Retrieval-Augmented Generation (RAG), which augments model generation with information retrieved from external knowledge sources~\cite{lewis2020retrieval}. Later on, ReAct interleaves reasoning with actions that allow language models to interact with external knowledge sources and environments and progressively acquire additional information during reasoning~\cite{yao2022react}. More recent work has brought relevant mechanisms together under the broader concept of context engineering, encompassing context retrieval and generation, context processing, and context management~\cite{mei2025survey}. Engineering appropriate context is essential for AI systems operating in domain-specific applications. In the legal domain, LegalBench-RAG investigates the retrieval of precise legal text for downstream legal reasoning, highlighting the importance of retrieving highly relevant passages rather than providing large amounts of legal documentation~\cite{pipitone2024legalbench}. In healthcare, medical RAG systems can be sensitive to irrelevant or unhelpful context~\cite{sohn2025rationale}. Our work addresses a complementary design challenge. We consider domain constraints as first-class drivers for determining what context should be identified, structured, represented, and operationalised for AI systems.

\subsection{Domain-Driven and Architecture-Driven Design}

Domain-Driven Design (DDD) emphasises that software design should be grounded in a deep understanding of the domain in which the system operates. DDD places the domain model at the centre of the development process~\cite{evans2004domain}. The model is developed collaboratively with domain experts through a ubiquitous language, providing a shared vocabulary for describing domain concepts, relationships, rules, and activities. DDD also introduces bounded contexts, which define explicit boundaries within which a particular domain model and its terminology remain consistent. A recent empirical synthesis confirms that these concepts support the translation of complex real-world domains into models that can guide software design ~\cite{ozkan2025domain}.

Software architecture design complements this domain-oriented perspective by considering a broader set of architectural drivers that shape a system. Architectural decisions are influenced by quality attributes and constraints imposed by the technical and organisational environment~\cite{bass2021software}. The Attribute-Driven Design (ADD) method operationalises this perspective through a design process in which quality-attribute requirements and other architectural drivers guide the selection of architectural tactics, patterns, and design decisions~\cite{wojcik_2006}.

\begin{figure}[t]
\centering\includegraphics[width=0.45\linewidth]{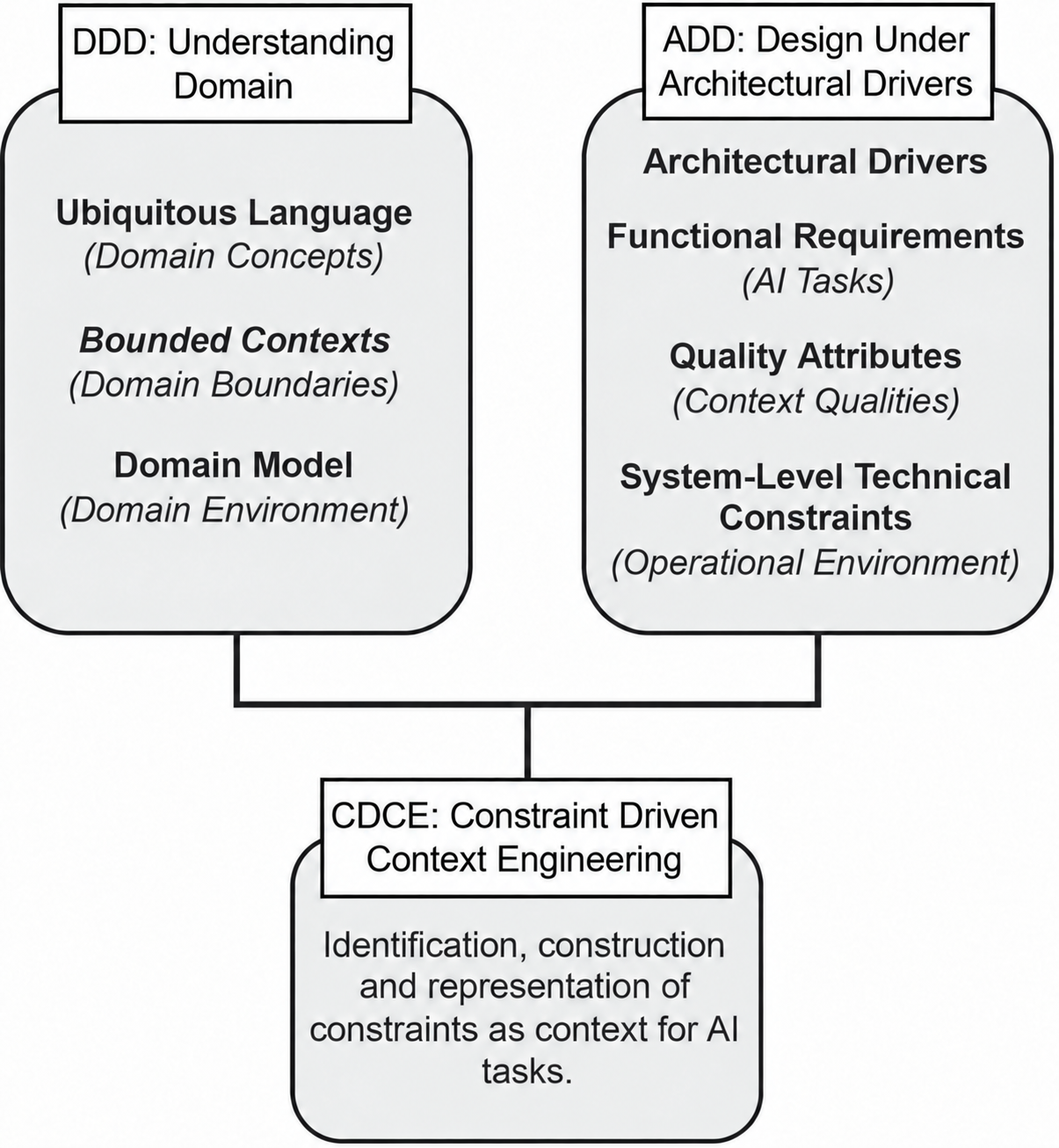}
\caption{Conceptual foundations of Constraint-Driven Context Engineering (CDCE). CDCE draws on Domain-Driven Design (DDD) to understand the domain and its boundaries, and on Attribute-Driven Design (ADD) to identify architectural drivers for AI tasks.}
\Description{A diagram showing Constraint-Driven Context Engineering as drawing on Domain-Driven Design and Attribute-Driven Design. DDD contributes ubiquitous language, bounded contexts, and a domain model. ADD contributes architectural drivers, functional requirements for AI tasks, quality attributes for context qualities, and system-level technical constraints for the operational environment.}
\label{fig:cdce-foundations}
\end{figure}

CDCE combines these two perspectives by connecting domain understanding with architectural reasoning: the former establishes what is meaningful in the domain (identify and model constraints), while the latter provides a systematic way to translate design drivers into architectural decisions (selection of context assets and representations). As shown in Figure\ref{fig:cdce-foundations}, we adopt DDD to identify and model constraints in the target domain. ADD treats these constraints as architectural drivers and translates the resulting constraints into context-design decisions, including the selection of assets and representations.

\subsection{Human Knowledge and Human-in-the-Loop}

Human knowledge has been recognised as an important source of information for intelligent systems. A fundamental distinction exists between explicit knowledge, which can be articulated and documented, and tacit knowledge, which is embedded in human experience, skills, and judgement. The concept of tacit knowledge was introduced based on the observation that people can possess knowledge that they cannot fully articulate~\cite{polanyi1967tacit}. Externalisation was later conceptualised as a process through which tacit knowledge is articulated into explicit concepts~\cite{nonaka1994dynamic}. More recently, Human-in-the-Loop (HITL) has incorporated human knowledge and oversight into different stages of AI system development and operation~\cite{mosqueira2023human}. Human involvement is particularly important in domain-specific applications, such as healthcare~\cite{olawade2026human} and education~\cite{memarian2024human}, where AI outputs need to be interpreted and evaluated within established professional practices. The professional knowledge and judgement that come from domain experts may contain implicit constraints that need to be externalised before being used by AI systems. CDCE seeks to externalise relevant implicit constraints from human knowledge and professional practice.

\section{Constraints}

CDCE treats domain constraints as first-class drivers for context engineering. We define a constraint as an environmental signal that guides, restricts, or evaluates AI-system behaviour and outcomes within a domain. 

\subsection{Characterisation of Constraints}

Domain constraints are heterogeneous and may differ in how they are expressed, when they become relevant, and how strictly they should be satisfied. We characterise constraints in terms of explicitness, temporality, and binding strength. 


\textbf{Explicitness} describes whether a constraint has been articulated and is directly identifiable. \textit{Explicit constraints} are expressed in identifiable sources, such as regulations, organisational policies, technical specifications, guidelines, or documented assessment criteria. \textit{Implicit constraints} are embedded in organisational practices, professional judgement, or tacit knowledge and may not be directly documented. 
Explicitness affects how constraints are identified and captured during context engineering. For example, as illustrated by the educational assessment case in Section~\ref{sec:education}, curriculum requirements and marking guidelines represent explicit constraints that are formally documented and directly identifiable. In contrast, teachers may rely on implicit professional judgement, such as expectations about what constitutes sufficient evidence of conceptual understanding. 

\textbf{Temporality} describes when a constraint is established or becomes relevant to an AI task. \textit{Predefined constraints} are known or established before the AI task begins, such as existing regulations, policies, or assessment requirements. \textit{Dynamic constraints} arise, change, or become relevant during operation as the state of the operational environment evolves. They may, for example, result from runtime conditions, ongoing interactions, or newly available information. Temporality affects when constraints need to be acquired, updated, and made available to the AI system. For example, as illustrated in Section~\ref{sec:finance}, the Audit Office's existing assessment criteria constituted predefined constraints, while subsequent interaction between the AI model and human auditors exposed inconsistencies and led to the refinement of these criteria. This illustrates how constraints may evolve dynamically during operation. 

\textbf{Binding strength} describes how strictly a constraint should govern an AI task and its outcomes. \textit{Hard constraints} represent mandatory conditions that must be satisfied, whereas \textit{soft constraints} provide guidance while allowing flexibility in their application, such as admitting exceptions, allowing discretion in how they are satisfied, or balancing them against other considerations. This distinction is particularly important when determining how a constraint should be operationalised: some constraints may require strict compliance, while others are more appropriately incorporated as contextual guidance for domain tasks. For example, as illustrated in Section~\ref{sec:healthcare}, applicable clinical practice requirements may constitute hard constraints that must be satisfied in care-plan generation, whereas recommendations from optional clinical handbooks or supporting literature may constitute soft constraints that guide AI behaviour without being mandatory. 

The three characteristics are conceptually distinct and constraints may exhibit different combinations of them. 
For example, a statutory privacy requirement may be explicit, predefined, and hard, whereas a preference expressed during an interaction may be explicit, dynamic, and soft. Together, these characteristics provide a systematic way to characterise heterogeneous domain constraints and inform their treatment in the subsequent context engineering process. 

\subsection{Types of Constraints}

We distinguish four complementary categories of constraints: \textbf{technical}, \textbf{regulatory}, \textbf{institutional}, and \textbf{normative}. These categories provide practical design abstractions rather than a fixed or exhaustive taxonomy; domains may instantiate them differently and introduce additional domain-specific constraints. 


\textbf{Technical constraints} arise from the capabilities and limitations of the underlying AI infrastructure and define the operational boundaries. Representative constraints include bounded reasoning capacity, state continuity, and other operational constraints. For example, the finite context window of an LLM limits the amount of information available during a single inference pass, driving context composition mechanisms such as selective retrieval, prioritisation, summarisation, and incremental context loading. 
Other operational constraints, such as latency, computational cost, deployment infrastructure, and domain-specific hardware, may further determine how context can be constructed and provided at runtime.


\textbf{Regulatory constraints} arise from laws and regulations that govern how AI systems operate within a domain. 
Regulatory constraints vary across jurisdictions and industries. Representative regulatory constraints include data access and use constraints, which determine what information may be accessed and processed; data retention and disclosure constraints, which govern how information may be stored and shared; and compliance and accountability constraints, which require AI systems and processes to comply with applicable regulations and standards. These constraints determine what information an AI system is permitted to access, process, retain, and disclose, as well as the obligations that AI outcomes must satisfy. 


\textbf{Institutional constraints} arise from the organisational environment in which AI systems operate. They include organisational structures, roles and responsibilities, operational procedures, workflow rules, and approved knowledge sources. While regulations define what an organisation must comply with, institutional constraints determine how work is performed within a particular organisation. 
Context engineering should therefore support role-aware context composition and workflow-aware behaviour, ensuring that recommendations are appropriate for both the task and the organisational setting in which they are made.


\textbf{Normative constraints} arise from professional judgement, accepted practices, and evaluation standards within a domain. 
Experienced professionals frequently rely on tacit knowledge, previous cases, and domain expertise when making decisions. Teacher judgement in educational assessment, clinical judgement in healthcare, and auditor judgement in financial auditing are all examples of normative constraints that influence AI behaviour.
Since professional judgement evolves over time, context engineering should also support continuous refinement through expert review, overrides, and feedback, allowing normative constraints to be progressively clarified and incorporated into future reasoning.

\subsection{Incorporating Constraints into Context Management Harnesses}

In this subsection, we show how the identified constraints can be incorporated into the context engineering within an existing harness framework. We adopt the notation of the DeepSeek harness~\cite{shi2026spatiotemporal} and represent a unified context management system as $\Gamma_{\infty}$. A particular runtime context state, denoted by $\gamma_t \in \Gamma_{\infty}$, includes session history, memory, data access, tools, retrieval services, etc. The formulation is implementation-agnostic. Constraints may be realised through different components and compositions in different harness frameworks. 

Let the context management system $\mathcal{H}$ be composed of components $\{c_i\}$, i.e.,
{\small
\[ \mathcal H =
\{c_1,\ldots,c_m\},
\qquad
c_i=(d_i,p_i,e_i),
\]
}
where \(d_i\) denotes the dependencies required by component \(c_i\),
\(p_i\) denotes the capabilities it provides, and \(e_i\) denotes its
reversible effects.

The four types of constraints at iteration \(t\) of the agent's problem solving are as follows:
{\small
\[
\mathcal K^{(t)}
=
\mathcal K_{\mathrm{reg}}
\cup
\mathcal K_{\mathrm{inst}}^{(t)}
\cup
\mathcal K_{\mathrm{norm}}
\cup
\mathcal K_{\mathrm{tech}},
\]
}
Different type of constraints may evolve at different paces, and we observed in the case studies that only institutional constraints are flexible for human experts to change. For the purpose of this formulation, we assume that only the institutional constraint set evolves across iterations as new evidence becomes available, while the regulatory, normative, and technical constraint sets remain fixed. 

Let $q$ denote the user query, and let $X_{q,t}^{\mathrm{adm}}$ denote the admissible evidence context for query $q$. The constraint-driven context engineering process follows the lifecycle below:
{\tiny
\[
\begin{aligned}
	\mathcal H
	&=
	\{c_1,\ldots,c_m\}
	\\
	&\Downarrow\
	\text{\small components interact through the unified runtime context}
	\\[2mm]
	\gamma_t
	&\in
	\Gamma_{\infty}
	\\
	&\Downarrow\
	\text{\small exposes available retrieval, memory, tools, and policies}
	\\[2mm]
	X_{q,t}^{\mathrm{adm}}
	&=
	\pi_{\phi}
	\left(
	q,\gamma_t, \mathcal{D};
	\mathcal K^{(t)}
	\right)
	\quad
	\text{s.t. }
	X_{q,t}^{\mathrm{adm}}
	\models
	\mathcal K^{(t)}
	\\
	&\Downarrow\
	\text{\small conditions model inference}
	\\[2mm]
	Y_t
	&\sim
	P_{\theta}
	\left(
	\cdot
	\mid
	q,X_{q,t}^{\mathrm{adm}},h
	\right)
	\\
	&\Downarrow\
	\text{\small enforces current decision constraints}
	\\[2mm]
	Z_t,F_t
	&=
	\operatorname{AutoVerify}
	\left(
	Y_t,
	X_{q,t}^{\mathrm{adm}},
	\mathcal K^{(t)}
	\right)
	\\
	&\Downarrow\
	\text{\small evaluates the utility of verified output}
	\\[2mm]
	V_t
	&=
	\mathbb E
	\left[
	U(Z_t)
	\right]
	\\	
	&\Downarrow\
	\text{\small human review may reveal a previously unmodelled inconsistency}
	\\[2mm]
	k_{t}^*
	&=
	\operatorname{HumanCheck}
	\left(
	Y_{t},
	Z_{t},
	F_{t},
	V_{t}
	\right)
	\\
	&\Downarrow\
	\text{\small incorporates the validated refinement}
	\\[2mm]
	\mathcal K_{\mathrm{inst}}^{(t+1)}
	&=
	\mathcal K_{\mathrm{inst}}^{(t)}
	\cup
	\{k_{t}^{*}\}
	\\
	&\circlearrowleft\
	\text{\small reruns the same input under the refined institutional constraints}.
	\\
	\end{aligned}
\]
}

\noindent where $\pi_{\phi}$ is the context management policy, $P_\theta$ is the LLM deployed, $\mathcal{D}$ denotes data sources, $h$ is the interaction history, $Y_t$ is the candidate model output, $Z_t$ is the verified output, $F_t$ is the verification feedback, $V_t$ is the expected utility of output $Z_t$. 
In this flow, the context admission and AutoVerify components check the agent output using known constraints, while the HumanCheck mechanism updates institutional constraints based on observed inconsistencies due to unmodelled factors revealed by the agent output. The agent output is evaluated by a utility function $U$ that measures the impact of taking the agent output as the decision. The three case studies illustrate different configurations of these operations across three domains.

\section{Constraint-Driven Context Engineering (CDCE)}

\begin{figure}[t]
\centering\includegraphics[width=0.85\linewidth]{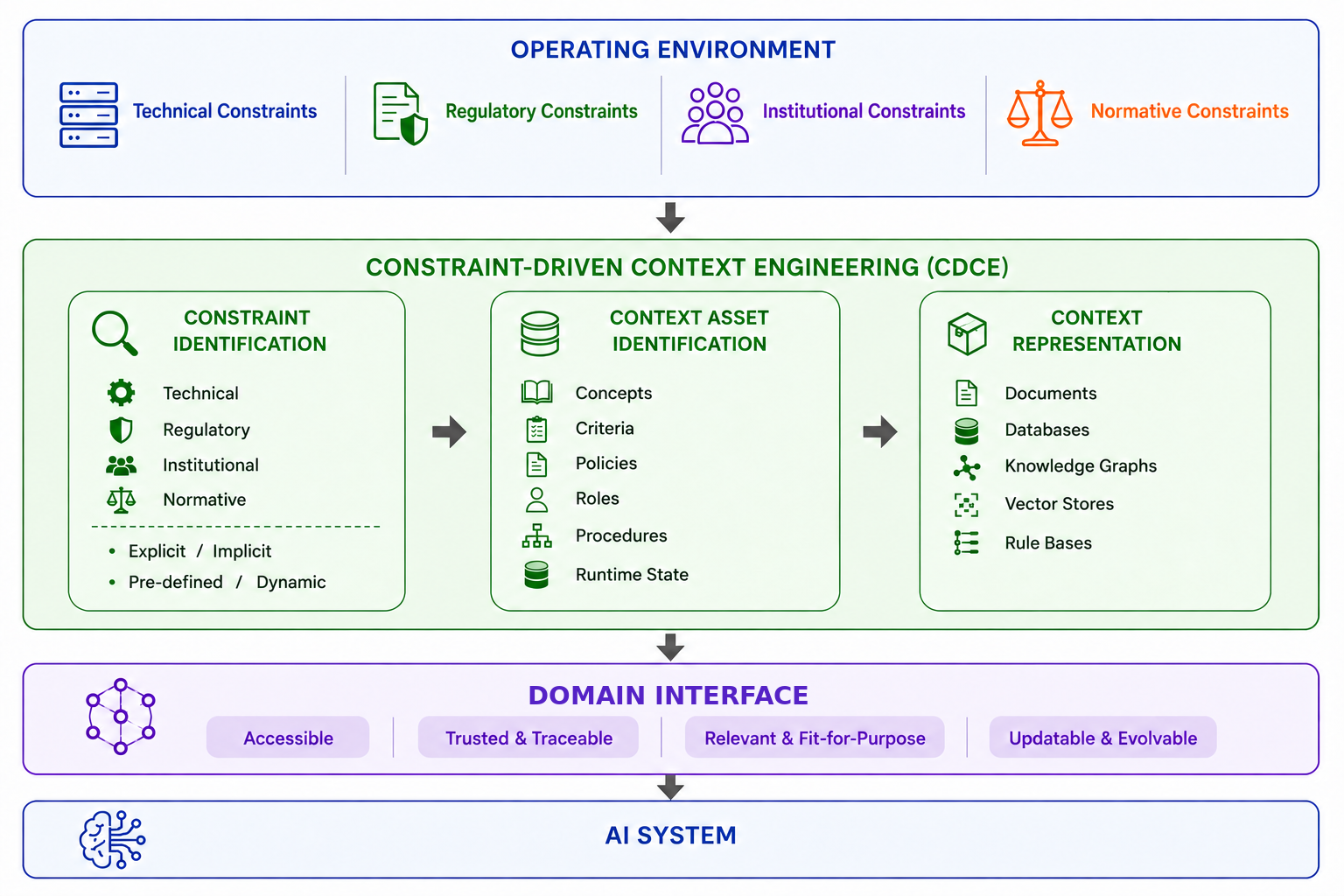}
    \caption{Overview of the proposed Constraint-Driven Context Engineering (CDCE).}
    \Description{Overview of the proposed Constraint-driven Context Engineering (CDCE).}
    \label{fig:overview}
\end{figure}

Figure \ref{fig:overview} presents the proposed Constraint-Driven Context Engineering approach. The framework views context engineering as a design process that transforms information from the operational environment into structured and machine-accessible context for domain-specific AI tasks. The proposed approach systematically identifies the constraints that govern AI behaviour within the operational environment, further identifies the relevant context assets and engineers them into appropriate representations that can be consumed by AI systems. The operational environment provides the input to CDCE, while the primary output is a domain interface, which is the engineered combination of context assets and representations.

The framework consists of three stages. The process begins by identifying the constraints that govern AI behaviour within the operational environment. 
These constraints determine not only what an AI system should know, but also how it should reason, what obligations it must satisfy, and how its outputs should be evaluated.

Once the required context assets have been identified, they are represented using appropriate data structures and technologies that support efficient access. Different types of information may be represented as documents, structured databases, knowledge graphs, vector stores, or rule bases according to their characteristics and intended use. The framework separates what information is required from how that information is represented, allowing different representation technologies to be selected without changing the underlying reasoning model.

These engineered context assets and their representations form the domain interface, an engineered boundary between an AI system and its operational environment. 
The AI system remains responsible for performing its tasks and generating responses, while CDCE governs the quality, scope, and structure of the context that forms its domain interface.

\subsection{Constraint Identification}

This stage begins with developing a shared understanding of the domain through close collaboration with domain experts~\cite{evans2004domain}. DDD emphasises the use of a Ubiquitous Language shared by developers and domain experts to describe domain concepts and activities consistently. We use this principle to establish a common vocabulary for describing the concepts, actors, activities, rules, and judgements involved in the target AI task. The use of domain language is particularly important for AI tasks, as the AI task and its contextual information are ultimately expressed and interpreted largely through natural language.

DDD also recognises that domain concepts and models may have different meanings in different parts of a complex domain and uses Bounded Contexts to establish explicit boundaries within which a particular model and its terminology remain consistent~\cite{evans2004domain, ozkan2025domain}. We adopt this principle to define the scope of constraint identification. Constraint identification is therefore task-oriented: the objective is not to model all aspects of the operational environment, but to determine which environmental conditions guide, restrict, or evaluate AI behaviour within the bounded context of the target AI task. Relevant constraints may be identified through domain artefacts such as regulations, organisational procedures, professional guidelines, and technical and operational documentation, or elicited through collaboration with domain experts where they are embedded in established practices and professional judgement. The output of this stage is a task-specific set of constraints that defines the conditions under which the target AI task should take place. 

\subsection{Context Asset Identification}

Once the constraints have been identified, the next stage is to determine the context assets required to represent or operationalise those constraints for AI tasks. Context assets are the information artefacts that provide the necessary context within the identified constraints. Depending on the domain, they may include domain concepts, judgement criteria, regulations and policies, organisational roles, operational procedures, runtime state, or other domain-specific information.

A single constraint may require multiple context assets to operationalise it, while the same context asset may support reasoning under multiple constraints. For example, a regulatory constraint may require not only the regulation itself, but also organisational policies, role information, operational procedures, and provenance information to determine how the requirement should be applied within a particular AI task. The identified constraints and corresponding context assets form a domain sub-model that captures the information required for AI tasks within the operational environment. 

\subsection{Context Representation}

Different types of context assets possess different characteristics and therefore require different representation strategies. For example, a policy document differs significantly from runtime state, organisational workflows, or decision criteria in terms of structure, update frequency, semantic relationships, and retrieval requirements. The objective of this stage is to determine appropriate representations that enable diverse context assets to be effectively accessed and used by AI systems.

This stage applies ADD~\cite{bass2021software,wojcik_2006}, which treats functional requirements, quality attributes, and design constraints as architectural drivers that guide architectural design decisions. We adapt this principle to design context representation. In CDCE, the AI task corresponds to the functional requirement, context qualities capture the quality attributes (non-functional requirements) required for the AI task, and system-level technical constraints define the operational conditions within which context must be provided. 

Following the ADD principle, 
context representation begins by analysing each context asset in relation to these drivers. The AI task determines how an asset will be used; for example, whether it needs to support semantic retrieval, rule evaluation, relationship traversal, or direct human inspection. Context qualities specify properties expected of the representation, such as relevance, explainability, traceability, consistency, and timeliness. System-level technical constraints, such as context-window limitations, latency, infrastructure, and operational cost, further restrict the feasible representation choices. The characteristics of the context asset itself, including human readability, semantic similarity, dependencies, explicit rules, structured data requirements, provenance, and update frequency, provide additional inputs to the representation decision.

The mapping between context assets and context representations is inherently many-to-many. A policy, for example, may be represented simultaneously as a document for human interpretation, a vector embedding for semantic retrieval, and a rule base for automated compliance checking. Conversely, a knowledge graph may integrate concepts, policies, roles, and procedures originating from multiple context assets. 
Separating context assets from context representations provides architectural flexibility. It allows the same context asset to be represented using multiple complementary structures while enabling representation technologies to evolve independently of the underlying context assets. 

\section{Comparative Multiple-Case Study Methodology}

We conducted a comparative multiple-case study~\cite{yin2017case} to examine the applicability of the CDCE approach across real-world AI systems in different domains. Case study research is appropriate for examining a contemporary phenomenon within its real-world context. In this study, the phenomenon of interest is the construction of a domain interface between an AI system and its operational environment.

\subsection{Case Selection}

The cases were selected to cover diverse application domains and AI tasks with different characteristics. They comprise educational assessment, healthcare decision support, and financial-distress prediction, and were conducted through commercial projects with Australian industry partners and a public-sector collaboration partner. All three projects emerged from a recent pattern of industry and public sector engagement. Partners typically approached the research team with a functioning but relatively generic, or vanilla, AI solution. The central challenge was not to build an AI system from scratch, but to improve the quality and domain appropriateness of an existing solution. In these settings, the limiting factor was often not the underlying model capability, but the quality, scope, and structure of the context available to the system.

The three cases were selected because they illustrate how context engineering becomes necessary when generic AI systems must operate within distinct domain conditions. Although the domains and AI tasks differ, each case required the identification and operationalisation of domain-specific requirements, organisational processes, data, and professional judgement to improve the suitability of the AI system for its intended setting. Educational assessment involves curriculum requirements and teacher judgement; healthcare decision support involves clinical requirements, patient-specific context, and clinician oversight; and financial-distress prediction involves heterogeneous operational data, historical assessments, and evolving auditor judgement. Together, the cases enable comparison of how CDCE is applied across contrasting constraint-rich domains.

\subsection{Case-Study Protocol and Analytical Focus}

Following Yin’s multiple-case design, we use a common case-study protocol to analyse the three cases consistently and support cross-case synthesis~\cite{yin2017case}. The protocol is derived from the CDCE design process. It provides a common analytical lens for examining how CDCE was applied and adapted across the cases. Each case is presented using the three CDCE design stages: constraint identification, context asset identification, and context representation design. Each case then concludes with an observation that captures the distinctive insight arising from its application. This common structure enables comparison across cases while preserving the particular characteristics of each domain, AI task, and partner context.

The cases are not expected to instantiate CDCE in an identical manner. In practice, context engineering is iterative and case-specific. A constraint may be identified before an AI system is deployed, or it may become visible only when a preliminary system output is reviewed by domain experts. Similarly, a context asset may be directly available in an existing document or data source, constructed from multiple artefacts, or constructed by externalising professional judgement. The selected representations and associated operations also depend on the task, available infrastructure and context characteristics.

The analysis examines both the common CDCE design logic and the flexibility with which it can be applied. In the cross-case discussion, we consider which elements recur across the cases, how different constraint characteristics lead to different context-engineering decisions and refine domain interface. 

\section{Case Study I: Educational Assessment}
\label{sec:education}

Educational assessment requires teachers to evaluate student responses against curriculum expectations while exercising professional judgement. Unlike factual question answering, assessment requires reasoning over multiple sources of contextual information, including syllabus documents, learning outcomes, marking guidelines, cognitive demand, and established assessment practice. The objective is not simply to determine whether an answer is correct, but whether it satisfies the intent of the curriculum and the expectations of experienced assessors.

The first case study is based on a commercial project with Studitory\footnote{\url{https://www.studitory.app/}}, an online learning platform developed to support NSW Higher School Certificate (HSC) examination preparation~\cite{xu2026llm}. The platform served more than 5,700 registered students at the time of the project and deployed an LLM-based pipeline for automated marking and personalised feedback. The pipeline was designed for exam preparation rather than final certification and grounds marking in authorised curriculum and assessment practices.

\subsection{Constraint Identification} 

Following the CDCE process, the first activity was to understand the educational assessment domain and identify the constraints required to support assessment. Constraint identification was informed by regular discussions with teachers and tutors (domain experts) involved in the project. The research team also visited schools and tutoring centres to understand how educators interpret assessment tasks, construct marking criteria, assign marks, and provide feedback in practice. These engagements informed the identification of both documented requirements and the professional judgement embedded in assessment practice. The marking task was shaped by authorised curriculum requirements, established assessment processes, professional marking judgement, and the technical conditions of the LLM-based pipeline.
 
Curriculum requirements and formal assessment principles are largely explicit and predefined, whereas some aspects of teacher judgement are implicit, including experience-based expectations concerning sufficient evidence, common misconceptions, and appropriate partial credit. Professional judgement is not uniformly implicit. Some established marking practices have already been codified in authorised assessment materials, while other aspects remain embedded in marker heuristics or institutional moderation practices. The constraints also differ in binding strength. Curriculum requirements establish relatively firm assessment boundaries, while professional judgement may be softer and permit legitimate variation in how student responses are interpreted. The marking pipeline primarily operationalised these constraints through contextual guidance and verification rather than relying on a single prompt.

\subsection{Context Asset Identification}

\textbf{Regulatory constraints} arise from the authorised curriculum and assessment requirements established by the NSW education department. For curriculum alignment, the pipeline uses authorised syllabus assets including \textit{learn about} statements, which specify conceptual and content knowledge; \textit{learn to} statements, which describe cognitive skills; and \textit{outcomes}, which describe assessable capabilities. The question itself and its maximum mark provide additional task-specific context.

Assessment reasoning also requires context assets that operationalise the meaning and expected quality of student performance. The NESA glossary of key words provides jurisdiction-specific interpretations of directive verbs such as \textit{analyse}, \textit{explain}, and \textit{evaluate}. Performance band descriptors provide authorised descriptions of different levels of performance, while marking guideline principles capture established assessment norms such as consistency, fairness, recognition of alternative correct approaches, and proportional allocation of marks.

\textbf{Institutional constraints} arise from established assessment processes and the operational workflow of the Studitory platform. Relevant context assets include question text and maximum marks, mappings between questions and syllabus elements, generated marking criteria, performance-level descriptors, verification results, and intermediate assessment artefacts. These assets capture how assessment activities are organised and how marking decisions are produced, recorded, and reviewed.

\textbf{Normative constraints} arise from professional marking practice and teacher judgement. Teachers interpret cognitive demand, decide what constitutes sufficient evidence, recognise alternative correct approaches, allocate partial credit, and evaluate the overall quality of a response. Some of this judgement is already articulated through performance band descriptors, verb definitions, marking principles, and historical marking practices. These artefacts were therefore identified as judgement-oriented context assets rather than leaving the LLM to rely on its internal knowledge. There are other judgements that are hard to capture, like the density of domain-specific terms used in a paragraph, which indicates how well a student understands a specific concept. 

Not all required context assets exist directly in source documents. The marking pipeline progressively constructs task-specific context assets from these authorised sources. For each question, relevant syllabus concepts, skills, and outcomes are identified and verified for coherence. These assets, together with the directive verb and maximum mark, are then used to construct question-specific marking criteria. The criteria specify expected behaviours and are subsequently calibrated into performance levels describing different degrees of demonstrated understanding.

In this case, curriculum requirements, directive-verb definitions, and performance descriptors primarily guide the construction of task-specific marking criteria. Assessment principles and outcome requirements provide additional criteria for verifying the selected curriculum context and the generated marking criteria. The marking pipeline therefore uses context assets both as contextual guidance and as a basis for verification. Collectively, these context assets form a domain sub-model for educational assessment. 

\subsection{Context Representation Design}

The context assets are represented differently according to their characteristics and role in assessment reasoning. Authorised syllabus statements are extracted from source documents and persisted as structured data. Vector representations support semantic matching between a question and relevant \textit{learn about} and \textit{learn to} statements, while relationships among selected concepts, skills, and outcomes are subsequently checked through a coherence-verification stage.

Different assets were therefore represented in different ways. Curriculum documents and marking guidelines remained as human-readable documents while also being indexed for semantic retrieval. Question metadata, syllabus mappings, concepts, and skills were represented as structured data to support efficient retrieval and reasoning. Generated marking criteria and intermediate assessment artefacts were persisted as structured objects linked to each assessment task, while provenance information was maintained throughout the marking process to support inspection and audit. The context representations are summarised below.

\begin{itemize}
    \item \textbf{Documents:} Curriculum documents, marking guidelines, performance band descriptors, marking guideline principles.
    \item \textbf{Vector Store:} Curriculum documents, marking guidelines.
    \item \textbf{Structured Data:} Question metadata, syllabus mappings, concepts and skills.
    \item \textbf{Structured Objects:} Generated marking criteria, intermediate assessment artefacts.
    \item \textbf{Key-value Store:} Runtime state, verification results.
    \item \textbf{Semi-Structured Logs:} Provenance information.
\end{itemize}

\subsection{Observation: Progressive Context Enrichment}

A key observation from the Studitory case is that the context required for marking does not exist as a fixed collection of documents to be retrieved at inference time. Instead, the reasoning context is progressively constructed and enriched through the assessment pipeline. The process begins with the assessment question and progressively constructs richer context assets by identifying relevant syllabus outcomes, concepts, skills, directive verbs, and cognitive demand. These artefacts are subsequently used to generate assessment criteria before evaluating student responses and producing personalised feedback. Each stage creates new context assets that become inputs to subsequent reasoning stages.

This progressive enrichment transforms raw educational resources into an increasingly structured domain interface. The case illustrates that context engineering is a constructive design process through which richer task-specific context is engineered between an AI system and its operational environment.

\section{Case Study II: Healthcare Decision Support}
\label{sec:healthcare}

The second case study is based on a commercial project conducted with an industrial partner in the healthcare domain. The project focuses on multidisciplinary team (MDT) care, which includes “clinicians from multiple disciplines who work together to deliver comprehensive care addressing as many of the patient’s health and other needs as possible”~\cite{acsqhc2020communicating}. In this use case, MDT mainly deals with chronic disease care. An MDT develops a care plan for a patient through meetings that often involve the following steps: preparation of relevant information before the health professionals meet, discussion of a specific case among health professionals based on their professional knowledge and experience, and decision-making regarding the care plan. The information-preparation stage requires the patient’s history to be extracted accurately from the health record and appropriately referenced to support care-planning decisions during the discussion. In addition, the final care plan produced after the MDT meeting should be grounded in evidence and comply with relevant medical guidelines.

\subsection{Constraint Identification}

To identify the constraints on context retrieval, detailed interviews were conducted with clinicians to understand their workflows and document the factors affecting their decision-making at each stage. By participating in MDT meetings, we gained an understanding of the external sources that clinicians draw on to support their reasoning and to personalise care plan items for a patient's specific circumstances. We also inferred how clinicians assess contextual relevance when constructing arguments and justifying their recommendations against medical guidelines.

In addition, we gained an understanding of the boundaries among clinicians participating in an MDT meeting, including their respective areas of expertise, their access to domain knowledge, and how that knowledge was activated during discussions. In general, the following constraints on context management have been identified:

\textbf{Institutional}: Important patient evidence must be included in the context when producing a care plan item. This requires the AI system to identify relevant paragraphs from an expanding document corpus, extract the relevant information, and connect it with other evidence to produce grounded care plan items.

\textbf{Regulatory}: Generated care plan items must comply with applicable medical guidelines. This requires the AI system to retrieve relevant medical knowledge with high recall and link applicable rules or recommendations to individual care plan items.

\textbf{Normative}: A treatment plan item must be supported by evidence from the patient's history and include the reasoning that connects the evidence to the proposed plan. This requires a post-generation validation mechanism to check the completeness and validity of the reasoning process.

Finally, as the treatment of chronic diseases is a long-term process, changes in the patient’s history and treatment plan form part of the context for developing the current care plan. Context management therefore needs to address the scalability challenge of an expanding longitudinal record while enabling accurate retrieval of relevant evidence over time.

\subsection{Context Asset Identification}

Regulatory constraints include two categories in this case: the first comprises guidelines regulating health practitioners in Australia, including applicable Australian clinical practice guidelines like relevant RACGP guidelines and NHMRC-issued or NHMRC-approved guidelines~\cite{racgp2026clinical,racgp2017standards, nhmrc2026guidelines}; the second contains other guidelines, optional handbooks and medical literature assisting clinicians' reasoning. Specific requirements within the former may constitute hard constraints where compliance is mandatory, whereas the latter generally provides soft guidance that can be balanced against patient-specific considerations. 

Context access for the clinician agent developed in this project is also subject to institutional constraints. The agent must operate within the existing MDT workflow. The information accessible to the agent may be fragmented because of access restrictions imposed by privacy policies, while the patient record may contain different levels of detail about the patient’s treatment as documented by different clinicians. Meanwhile, AI-based ambient clinical documentation tools are increasingly being incorporated into clinical workflows. For example, Heidi is an AI medical scribe that transcribes clinician–patient consultations and MDT discussions and generates structured clinical notes and related documentation~\cite{BRACKEN2026119}. Notes or care plans generated by such tools may represent information differently from the original clinical notes, potentially weakening signals that are important to clinicians for certain cases.

Normative constraints arise from clinicians’ judgement and experience. Clinicians from different disciplines may have different perspectives on the same patient. Even clinicians within the same discipline may have different views because of differences in the patients they have treated and the clinical experience they have accumulated. During MDT discussions, clinicians not only consult one another about their respective areas of expertise, but also challenge one another’s judgements and scrutinise one another’s reasoning.

These constraints require the context-management system to achieve high recall in retrieval and to identify potentially relevant information that can subsequently be validated by clinicians. They also require the system to have self-verification capabilities to determine whether each care plan item is adequately grounded in evidence.

To achieve the former, we used clinician-confirmed care plans as ground truth to guide the agent in generating care plans that minimise discrepancies from clinician-approved decisions. The generation process that minimises these discrepancies was then encoded as a skill in the agent harness. To achieve the latter, the clinician decision-support agent was required to generate fine-grained decision provenance, and an independent evaluator was introduced to assess the reasoning process for correctness and coverage.

\subsection{Context Representation Design}

As described above, the context contains patient records, past and current care plans, different types of medical guidelines
. Their representations are summarised below: 

\begin{itemize}
    \item \textbf{Structured data:} Patient health record.
    \item \textbf{Documents:} Past patient care plans; RACGP and NHMRC clinical practice guidelines; other medical guidelines and literature; patient treatment plan.
    \item \textbf{Semi-structured logs:} Care plan provenance.
\end{itemize}


These assets are organised in a verification-oriented manner. Every care plan item must declare its support status. Guideline-informed items are supported by evidence in the patient record and applicable medical-guideline rules, while directly documented care plan items may rely on evidence in the patient record alone. Items without declared support remain explicitly identifiable for review. The representation is illustrated as follows:
\vspace{1em}
\begin{center}
\begin{minipage}{0.95\linewidth}
\small
\setlength{\fboxsep}{6pt}
\noindent\fbox{%
\begin{minipage}{0.93\linewidth}
\textbf{Patient Evidence:} The patient has condition x.$\rightarrow$ \\
\textbf{Medical Rule:} For patients with x, clinicians should consider y.$\rightarrow$ \\
\textbf{Decision:} Rule y is applicable as a discussion consideration. $\rightarrow$ \\
\textbf{Care Plan Item:} Discuss y with the patient and treating practitioner. $\rightarrow$ \\
\textbf{Care plan wording}
\end{minipage}%
}
\end{minipage}
\end{center}
\vspace{1em}

This structure can be mapped to elements of Toulmin method~\cite{toulmin2003uses} as follows. The Toulmin method models an argument as a structured relationship among a claim, supporting data, warrant, backing, qualifier, and possible rebuttal, making explicit how evidence justifies a conclusion.

\noindent\textbf{Toulmin mapping:}

\begin{itemize}
    \item \textit{Patient evidence} $\rightarrow$ \textit{Data/Grounds};
    \item \textit{Medical-guideline knowledge rule} $\rightarrow$ \textit{Warrant or warrant source};
    \item \textit{Apply knowledge to make decision} $\rightarrow$ \textit{Application of the warrant to this patient context};
    \item \textit{Care plan item} $\rightarrow$ \textit{Claim}.
\end{itemize}


Importantly, Toulmin roles are assigned relative to a particular inference,
rather than being fixed properties of semantic elements. These roles may change across a multi-stage reasoning process. In particular, an intermediate clinical proposition or decision may be the claim of one inference and subsequently become a ground for the next inference.






\subsection{Observations: Skill Acquisition and Robust Care Plan Generation}

The main observation from this case is that verification-oriented design can surface inconsistencies in patient records and raise clinicians’ awareness of potential problems. In one case, the dietitian asked about a patient’s bowel cancer test result because the record showed that the result from a government-supplied bowel cancer screening kit was pending. However, the patient’s age was outside the range for which the government would automatically issue such a testing kit, based on clinical practice guidelines. The agent identified this inconsistency, and the clinician confirmed the discrepancy in the patient record.

Another observation is that the agent can learn from clinicians to personalise care plans for patients with certain conditions by retrieving specific knowledge from a broader range of handbooks. In one case, the agent was able to recommend vegetable preparation methods for further reducing potassium as part of the care plan. This recommendation was in line with advice from the dietitian, who understood the patient’s usual diet.

\section{Case Study III: Financial Distress Prediction}
\label{sec:finance}

Financial distress assessment in local government is a data-intensive reasoning task that requires auditors to interpret financial performance, liquidity, infrastructure condition, demographic characteristics, and other contextual indicators collectively. It is also a judgement-intensive task with interpretation evolving with organisational practices, available evidence, and auditors' understanding of financial sustainability.

This case study is based on the Risk IQ project conducted with the audit office of a local government to support the identification and analysis of local councils at risk of financial distress. The project combined data from the local governments (LG), internal data from the audit office's audit reports, and financial distress literature. The primary LG dataset covered 128 councils and approximately 572 year-prefixed indicators across multiple years. These indicators spanned financial performance ratios, infrastructure and asset measures, service expenditure, demographic and economic characteristics, development activity, workforce, rates and charges, and governance indicators. The literature-augmented datasets increased the available feature space to approximately 646 year-prefixed indicators. The assessment of financial sustainability depended on the interpretation of multiple indicators together with professional judgement, and assessment practices had evolved over time. 

\subsection{Constraint Identification}

The first activity was to understand how auditors assess financial distress and to identify the constraints that must guide this reasoning. This involved examining established assessment practices, available operational data, historical assessment decisions, and the professional judgement used to interpret financial sustainability. At a high level, the assessment task is constrained by financial reporting and auditing requirements, organisational assessment processes, and auditor judgement, together with the technical characteristics of the underlying longitudinal data. These constraints exhibit different characteristics. Some are explicit and predefined, such as financial indicators, benchmark calculations, and established assessment criteria. Others are implicit in professional judgement, including how auditors interpret combinations of financial and contextual signals and distinguish different forms of financial risk. Historical assessment decisions provide further evidence of how such judgement has been applied in practice, but do not necessarily constitute internally consistent ground truth.

The constraints may also evolve through interaction between modelling and auditor review. In this case, the initial assessment framework did not clearly distinguish short-term cash liquidity risk from longer-term financial sustainability risk. The first modelling iteration exposed inconsistencies in the existing assessments and prompted auditors to revisit this distinction. The resulting clarification and revised criteria 
then became explicit constraints for subsequent analysis. 

\subsection{Context Asset Identification}

\textbf{Regulatory constraints} are embedded in the financial reporting and auditing environment within which councils operate. Relevant context assets included published audit office tabling reports, financial benchmarks, and the underlying financial indicators used to assess council performance. 

\textbf{Institutional constraints} are reflected in existing assessment practices and established organisational processes. Relevant context assets included historical tabling assessments, dashboard data, existing assessment categories, benchmark calculation procedures, and the relationships between reported financial measures and their underlying data. Together, these assets captured how financial sustainability assessments were operationalised within the organisation.

\textbf{Normative constraints} arise from auditor judgement about what constitutes financial distress and how different financial signals should be interpreted together. Historical assessments were treated as context assets capturing previous professional judgement. The 2025 assessment, for example, distinguished risks associated with ongoing operating losses or other indicators, low levels of available cash, and small or declining populations. During the project, analysis of the initial prediction model also exposed inconsistencies in how the existing criteria were interpreted across councils and years. Through discussion and review, auditors clarified that the existing assessment had been combining short-term cash liquidity and longer-term financial sustainability. Revised criteria and additional data were subsequently provided to better represent these dimensions.

\textbf{Technical constraints} arise primarily from the structure and temporal characteristics of the available data. Raw indicators were organised as year-prefixed columns, labels were sparse and year-specific, and potentially useful signals were distributed across hundreds of indicators. Financial indicators also exhibited different levels of temporal persistence, meaning that the relevant context for predicting distress could depend on both the indicator and its historical lag. The system therefore needed mechanisms for temporal alignment, feature selection, provenance preservation, and reuse of the same reasoning structure across prediction years.

\subsection{Context Representation Design}

The next activity concerned how the identified context assets should be represented and composed for financial distress assessment. The original context assets existed in heterogeneous forms: annual Excel workbooks with time-series datasets, audit reports, literature, historical labels, and auditor judgement. Different assets therefore required different representations, while multiple context assets could be consolidated into the same representation.

An important decision concerned how temporal context should be represented. The original datasets represented observations using calendar-year-prefixed fields such as 2022 operating performance ratio, 2023 operating performance ratio, and 2024 operating performance ratio. Risk IQ introduced a relative-lag representation in which these observations could be represented relative to the prediction year as \(n-3\), \(n-2\), and \(n-1\). This separated the meaning of a historical relationship from a particular calendar year and allowed multiple target years to share a common temporal representation.

The large number of available indicators also created a context-selection problem. Rather than exposing all available features to the prediction process, Risk IQ used lag-aware causal discovery to identify candidate upstream covariates associated with the target assessment. The discovery process excluded future-year information and other label columns to prevent target leakage and identified significant historical relationships across feature families. From a context-engineering perspective, this process reduced a broad set of available environmental signals to a task-specific set of contextual information relevant to financial distress prediction. A summary of the context representation is as follows: 

\begin{itemize}
    \item \textbf{Documents:} Audit reports, financial sustainability criteria, domain literature.
    \item \textbf{Structured Data:} Financial indicators, dashboard data, benchmark thresholds, historical risk assessments.
    \item \textbf{Computation Structures:} Financial formulas, benchmark calculation procedures.
    \item \textbf{Relative-Lag Features:} Multi-year financial and contextual indicators, temporal relationships.
    \item \textbf{Rules:} Existing assessment criteria, revised assessment criteria, auditor interpretations.
    \item \textbf{Human Feedback:} Auditor review, revised interpretations, feedback on model outcomes.
    \item \textbf{Model Artefacts:} Feature-selection provenance, model state, intermediate analysis artefacts.
\end{itemize}

Data-driven context selection was complemented by domain-driven context. A literature review of financial distress identified 20 fundamental causal factors, including revenue inadequacy, structural operating deficits, asset burden, liquidity weakness, and socioeconomic disadvantage. These factors were mapped onto corresponding LG features and could be combined with statistically discovered covariates. Provenance was retained to distinguish features selected through causal discovery, identified from the literature, or supported by both. This enabled the reasoning environment to combine empirically discovered relationships with established domain knowledge rather than relying exclusively on either source.

The contextual information required by financial distress assessment may not exist explicitly before context engineering begins. Financial distress concepts were distributed across hundreds of financial and contextual indicators, annual workbooks, calculation structures, assessment reports, and historical decisions. The design process progressively transformed raw and heterogeneous domain information into representations suitable for AI use: operational observations became temporally aligned feature families; financial formulas became explicit computation structures; historical assessments became structured judgement signals; and literature-derived factors became mapped domain knowledge. 

\subsection{Observation: Human-AI Co-Evolution of Constraints}

Historical auditor assessments initially provided normative signals for model development. However, these signals were not immutable ground truth. The first modelling iteration exposed inconsistencies in the existing assessment criteria and required further examination by auditors. This process helped clarify the distinction between cash liquidity risk and longer-term financial sustainability risk and led to revised criteria and additional contextual data. Human feedback in this case changed the interface. Auditor review refined the concepts, criteria, and data used by subsequent models. The second iteration consequently operated with a richer and more clearly defined set of environmental signals, producing stronger alignment between model outcomes and auditor judgement.

This case illustrates that constraints may evolve through interaction between AI systems and domain experts. Existing assessment criteria initially served as predefined constraints, while ambiguities in their interpretation became explicit through model development and human review. Once articulated, these dynamically refined constraints became explicit, predefined constraints for subsequent iterations. The domain interface can therefore evolve as professional understanding and organisational practices are refined.

Normative constraints should not always be treated as static rules encoded before deployment. Interaction with AI systems can expose ambiguities or inconsistencies in existing human practices. Context engineering consequently supports the capture of human overrides, revised criteria, additional evidence, and the rationale for disagreement, allowing the domain interface to evolve together with professional understanding.

\section{Discussion}

The three case studies illustrate how CDCE manifests differently across domains while following the same design process. 

\subsection{Domain Interface Construction}

Across the three cases, context engineering involved constructing domain interfaces rather than retrieving domain knowledge. In educational assessment, context was progressively enriched from curriculum resources into task-specific concepts, skills, cognitive demand, marking criteria, and assessment artefacts. In healthcare decision support, patient evidence and medical knowledge were composed into structured relationships connecting evidence, applicable guidelines, decisions, and care-plan items. In financial distress prediction, relevant context was reconstructed from heterogeneous operational indicators, historical assessments, domain literature, and auditor judgement.

These cases demonstrate that context may need to be retrieved, constructed, composed, or reconstructed depending on the characteristics of the domain and task. Context engineering is a design activity that determines what domain conditions need to be made available to an AI system and how heterogeneous information should be transformed into a domain interface suitable for AI use.

\subsection{Constraints as Guidance, Enforcement, and Verification}

The cases also illustrate that constraints can play different roles within a domain interface. Some constraints primarily provide guidance to AI reasoning. Curriculum requirements and performance descriptors guide educational assessment, while clinical guidelines provide knowledge and recommendations relevant to care planning. Other constraints provide criteria against which generated artefacts can be independently verified. In educational assessment, generated marking criteria are checked against established assessment principles; in medical decision support, care-plan items are evaluated against their supporting patient evidence and applicable medical knowledge; and in Risk IQ, model outcomes and assessment criteria are reviewed by auditors.

Making a constraint available as contextual information is not always sufficient to ensure that it is satisfied. Depending on its binding strength and the consequences of violation, a constraint may need to guide AI behaviour, be enforced through system mechanisms, or provide criteria for independent verification. These roles are complementary rather than mutually exclusive: the same constraint may be represented as context to guide generation while also being independently verified or enforced where stronger assurance is required.

\subsection{Evolving Domain Interfaces}

The three cases further demonstrate different ways in which a domain interface can evolve. In Studitory, relatively stable curriculum and assessment constraints support the progressive construction of task-specific context during the marking process. In the healthcare case, clinical guidelines may remain relatively stable while patient state, available evidence, and care plans evolve over time. In Risk IQ, the constraints themselves evolved: the first modelling iteration exposed ambiguities in existing assessment criteria, and subsequent auditor review clarified the distinction between liquidity risk and longer-term financial sustainability, producing revised constraints for later iterations.

An operational environment may continuously produce new contextual information while its governing constraints remain stable; alternatively, interaction between AI systems and domain experts may reveal ambiguities that require the constraints themselves to be refined. Context and harness engineering should support both the maintenance of changing context and the evolution of the domain interface as professional knowledge, organisational practices, and environmental conditions change.

\section{Conclusion}

This paper introduced Constraint-Driven Context Engineering (CDCE), a design approach that treats technical, regulatory, institutional, and normative constraints as first-class drivers for engineering AI systems. CDCE systematically identifies and characterises domain constraints, determines the required context assets, and designs appropriate representations. Together, these engineered elements form a domain interface between an AI system and its operational environment, through which constraints can guide AI-system behaviour or inform complementary enforcement and verification mechanisms.

The three case studies demonstrate distinct aspects of constraint-driven context engineering: progressive context enrichment in educational assessment; explicit boundaries and verification in healthcare decision support; and context reconstruction and dynamic constraint refinement in financial-distress prediction. Together, they show that context engineering is a design activity for constructing and maintaining the domain interfaces through which AI systems operate in constraint-rich domains.

\section{Acknowledgments}

ChatGPT was used to assist with figure preparation and language editing. The authors reviewed and verified all generated content and remain responsible for the final manuscript.

\bibliographystyle{ACM-Reference-Format}
\bibliography{FSE2027}

@String{Computing = "Computing" }

@String{Springer = "Springer-Verlag" }

@article{Perry1992,
author = {Perry, Dewayne E. and Wolf, Alexander L.},
title = {Foundations for the study of software architecture},
year = {1992},
issue_date = {Oct. 1992},
publisher = {Association for Computing Machinery},
address = {New York, NY, USA},
volume = {17},
number = {4},
issn = {0163-5948},
url = {https://doi.org/10.1145/141874.141884},
doi = {10.1145/141874.141884},
journal = {SIGSOFT Softw. Eng. Notes},
month = oct,
pages = {40–52},
numpages = {13}
}

@book{bass2021software,
  author    = {Len Bass and Paul Clements and Rick Kazman},
  title     = {Software Architecture in Practice},
  edition   = {4},
  publisher = {Addison-Wesley Professional},
  year      = {2021},
}

@techreport{wojcik_2006,
author={Wojcik, Robert and Bachmann, Felix and Bass, Len and Clements, Paul and Merson, Paulo and Nord, Robert and Wood, William},
title={Attribute-Driven Design (ADD), Version 2.0},
month={Nov},
year={2006},
number={CMU/SEI-2006-TR-023},
institution={Software Engineering Institute, Carnegie Mellon University},
doi={10.1184/R1/6572066.v1},
url={https://doi.org/10.1184/R1/6572066.v1},
note={Accessed: 2026-Sep-18}
}

@book{evans2004domain,
  title={Domain-driven Design: Tackling Complexity in the Heart of Software},
  author={Evans, E.},
  isbn={9780321125217},
  lccn={30503310},
  url={https://books.google.com.au/books?id=xColAAPGubgC},
  year={2004},
  publisher={Addison-Wesley}
}

@article{brown2020language,
  title={Language models are few-shot learners},
  author={Brown, Tom and Mann, Benjamin and Ryder, Nick and Subbiah, Melanie and Kaplan, Jared D and Dhariwal, Prafulla and Neelakantan, Arvind and Shyam, Pranav and Sastry, Girish and Askell, Amanda and others},
  journal={Advances in neural information processing systems},
  volume={33},
  pages={1877--1901},
  year={2020}
}

@article{lewis2020retrieval,
  title={Retrieval-augmented generation for knowledge-intensive nlp tasks},
  author={Lewis, Patrick and Perez, Ethan and Piktus, Aleksandra and Petroni, Fabio and Karpukhin, Vladimir and Goyal, Naman and K{\"u}ttler, Heinrich and Lewis, Mike and Yih, Wen-tau and Rockt{\"a}schel, Tim and others},
  journal={Advances in neural information processing systems},
  volume={33},
  pages={9459--9474},
  year={2020}
}

@article{schick2023toolformer,
  title={Toolformer: Language models can teach themselves to use tools},
  author={Schick, Timo and Dwivedi-Yu, Jane and Dess{\`\i}, Roberto and Raileanu, Roberta and Lomeli, Maria and Hambro, Eric and Zettlemoyer, Luke and Cancedda, Nicola and Scialom, Thomas},
  journal={Advances in neural information processing systems},
  volume={36},
  pages={68539--68551},
  year={2023}
}

@article{wei2022chain,
  title={Chain-of-thought prompting elicits reasoning in large language models},
  author={Wei, Jason and Wang, Xuezhi and Schuurmans, Dale and Bosma, Maarten and Xia, Fei and Chi, Ed and Le, Quoc V and Zhou, Denny and others},
  journal={Advances in neural information processing systems},
  volume={35},
  pages={24824--24837},
  year={2022}
}

@article{yao2022react,
  title={React: Synergizing reasoning and acting in language models},
  author={Yao, Shunyu and Zhao, Jeffrey and Yu, Dian and Du, Nan and Shafran, Izhak and Narasimhan, Karthik and Cao, Yuan},
  journal={arXiv preprint arXiv:2210.03629},
  year={2022}
}

@article{mei2025survey,
  title={A survey of context engineering for large language models},
  author={Mei, Lingrui and Yao, Jiayu and Ge, Yuyao and Wang, Yiwei and Bi, Baolong and Cai, Yujun and Liu, Jiazhi and Li, Mingyu and Li, Zhong-Zhi and Zhang, Duzhen and others},
  journal={arXiv preprint arXiv:2507.13334},
  year={2025}
}

@article{pipitone2024legalbench,
  title={Legalbench-rag: A benchmark for retrieval-augmented generation in the legal domain},
  author={Pipitone, Nicholas and Alami, Ghita Houir},
  journal={arXiv preprint arXiv:2408.10343},
  year={2024}
}

@inproceedings{sohn2025rationale,
  title={Rationale-guided retrieval augmented generation for medical question answering},
  author={Sohn, Jiwoong and Park, Yein and Yoon, Chanwoong and Park, Sihyeon and Hwang, Hyeon and Sung, Mujeen and Kim, Hyunjae and Kang, Jaewoo},
  booktitle={Proceedings of the 2025 Conference of the Nations of the Americas Chapter of the Association for Computational Linguistics: Human Language Technologies (Volume 1: Long Papers)},
  pages={12739--12753},
  year={2025}
}

@book{polanyi1967tacit,
  title={The Tacit Dimension},
  author={Polanyi, M.},
  isbn={9780143414186},
  series={Anchor books},
  url={https://books.google.com.au/books?id=jwLXAAAAMAAJ},
  year={1967},
  publisher={Anchor Books}
}

@article{nonaka1994dynamic,
  author  = {Nonaka, Ikujiro},
  title   = {A Dynamic Theory of Organizational Knowledge Creation},
  journal = {Organization Science},
  volume  = {5},
  number  = {1},
  pages   = {14--37},
  year    = {1994},
  url     = {http://www.jstor.org/stable/2635068}
}

@article{mosqueira2023human,
  title={Human-in-the-loop machine learning: a state of the art: E. Mosqueira-Rey et al.},
  author={Mosqueira-Rey, Eduardo and Hern{\'a}ndez-Pereira, Elena and Alonso-R{\'\i}os, David and Bobes-Bascar{\'a}n, Jos{\'e} and Fern{\'a}ndez-Leal, {\'A}ngel},
  journal={Artificial Intelligence Review},
  volume={56},
  number={4},
  pages={3005--3054},
  year={2023},
  publisher={Springer}
}

@article{olawade2026human,
  title={Human in the loop artificial intelligence in healthcare: applications, outcomes, and implementation challenges},
  author={Olawade, David B and Plabon, Shamiul Bashir and Ojo, Adeyinka and Ogunbona, Muyiwa Ademola and Makanjuola, Babajide David and Olasilola, Rosemary},
  journal={International Journal of Medical Informatics},
  pages={106362},
  year={2026},
  publisher={Elsevier}
}

@article{memarian2024human,
  title={Human-in-the-loop in artificial intelligence in education: A review and entity-relationship (ER) analysis},
  author={Memarian, Bahar and Doleck, Tenzin},
  journal={Computers in Human Behavior: Artificial Humans},
  volume={2},
  number={1},
  pages={100053},
  year={2024},
  publisher={Elsevier}
}

@article{ozkan2025domain,
  title={Domain-Driven Design in software development: A systematic literature review on implementation, challenges, and effectiveness},
  author={{\"O}zkan, Ozan and Babur, {\"O}nder and van den Brand, Mark},
  journal={Journal of Systems and Software},
  volume={230},
  pages={112537},
  year={2025},
  publisher={Elsevier}
}

@article{xu2026llm,
  title={LLM-as-Judge in Education: A Curriculum-Grounded Marking Pipeline},
  author={Xu, Xiwei and Wang, Chen and Jiang, Jacky and Yang, Phil and Fu, Qian and Dhall, Mohan and Zhang, Wenjie and Zhu, Liming},
  journal={arXiv preprint arXiv:2606.17507},
  year={2026}
}

@misc{acsqhc2020communicating,
  author = {{Australian Commission on Safety and Quality in Health Care}},
  title  = {Communicating for Safety: Improving Clinical Communication, Collaboration and Teamwork in Australian Health Services},
  year   = {2020},
  address = {Sydney}
}

@misc{racgp2026clinical,
  author       = {{Royal Australian College of General Practitioners (RACGP)}},
  title        = {Clinical Guidelines},
  year         = {2026},
  address      = {Melbourne},
  publisher    = {RACGP},
  url          = {https://www.racgp.org.au/clinical-resources/clinical-guidelines},
  urldate      = {2026-09-13}
}

@misc{racgp2017standards,
  author       = {{Royal Australian College of General Practitioners (RACGP)}},
  title        = {The RACGP's Standards for General Practices (5th Edition): A Benchmark for Quality Care and Risk Management in Australian General Practices},
  year         = {2017},
  url          = {https://www.racgp.org.au/running-a-practice/practice-standards/standards-5th-edition/standards-for-general-practices-5th-ed/}
}

@misc{nhmrc2026guidelines,
  author       = {{National Health and Medical Research Council (NHMRC)}},
  title        = {National Health and Medical Research Council (NHMRC) Guidelines},
  year         = {2026},
  url          = {https://www.nhmrc.gov.au/guidelines},
  urldate      = {2026-09-13}
}

@article{BRACKEN2026119,
title = {Ambient AI reduces documentation time and enhances quality in a simulated inpatient setting},
journal = {The Surgeon},
volume = {24},
number = {2},
pages = {119-125},
year = {2026},
issn = {1479-666X},
doi = {https://doi.org/10.1016/j.surge.2025.10.008},
url = {https://www.sciencedirect.com/science/article/pii/S1479666X25001544},
author = {Aisling Bracken and Anita Rose Babu and Seán Whelehan and Khalid Merghani and Eoin Sheehan and Iain Feeley}
}

@book{toulmin2003uses,
  title={The Uses of Argument},
  author={Toulmin, S.E.},
  isbn={9780521534833},
  lccn={2003043502},
  url={https://books.google.com.au/books?id=8UYgegaB1S0C},
  year={2003},
  publisher={Cambridge University Press}
}

@book{yin2017case,
  title={Case Study Research and Applications: Design and Methods},
  author={Yin, R.K.},
  isbn={9781506336183},
  url={https://books.google.com.au/books?id=6DwmDwAAQBAJ},
  year={2017},
  publisher={SAGE Publications}
}

@article{shi2026spatiotemporal,
	title   = {A Programming Paradigm for Spatiotemporal Composability},
	author  = {Shi, Yifan and Zhang, Wei and Cui, Tianyi},
	journal = {arXiv preprint arXiv:2608.25512},
	year    = {2026},
	doi     = {10.48550/arXiv.2608.25512}
}

\end{document}